\documentclass[reprint,superscriptaddress,
amsmath, amssymb,aps, prd,notitlepage,longbibliography,floatfix,
nofootinbib 
]{revtex4-1}

\usepackage{tensor}     
\usepackage{graphicx}   
\usepackage{subfigure}
\usepackage[
colorlinks=true,        
citecolor=blue,         
linkcolor=blue,         
urlcolor=blue           
]{hyperref}             
\usepackage{bm}         
\usepackage{xcolor}     
\usepackage{lipsum}
\usepackage{color}      
\usepackage[utf8]{inputenc} 
\usepackage[section]{placeins} 
\usepackage{multirow}
\usepackage[title]{appendix}
\usepackage{simpler-wick}
\usepackage{float}
\usepackage[normalem]{ulem}
\newcommand{\nc}{\newcommand*}

\nc{\al}{\alpha}
\nc{\s}{\sigma}
\nc{\kp}{\kappa}
\nc{\dt}{\delta}
\nc{\Dt}{\Delta}
\nc{\Ld}{\Lambda}
\nc{\p}{\partial}
\nc{\Gm}{\Gamma}
\nc{\om}{\omega}
\nc{\Om}{\Omega}
\nc{\rd}{\mathrm{d}}
\def\({\left(}
\def\){\right)}
\def\[{\left[}
\def\]{\right]}
\def\e{\begin{equation}}
\def\q{\end{equation}}
\def\be{\begin{equation}}
\def\ee{\end{equation}}
\def\m{\begin{eqnarray}}
\def\n{\end{eqnarray}}
\def\beq{\begin{eqnarray}}
\def\eeq{\end{eqnarray}}
\nc{\Eq}[1]{Eq.~\eqref{#1}}     
\nc{\Fig}[1]{Fig.~\ref{#1}}     
\nc{\Table}[1]{Table~\ref{#1}}  
\nc{\Sec}[1]{Sec.~\ref{#1}}     
\nc{\Msun}{M_\odot}             
\nc{\Ogw}{\Omega_{\mathrm{GW}}}
\nc{\gpcyr}{\mathrm{Gpc}^{-3}\,\mathrm{yr}^{-1}}
\nc{\lvc}{LIGO/Virgo} 
\nc{\SNR}{\mathrm{SNR}} 
\nc{\rhoGW}{\rho_{\mathrm{GW}}}
\nc{\vd}{\vec{d}}
\nc{\av}[1]{\langle #1 \rangle} 
\nc{\km}{\mathrm{km}}
\nc{\Mpc}{\mathrm{Mpc}}
\nc{\Tobs}{T_{\mathrm{obs}}}
\nc{\fyr}{f_{\mathrm{yr}}}
\nc{\hii}{\rm H~{\textsc {ii}}}
\nc{\addref}{[\textcolor{red}{add ref}] } 
\nc{\eg}{\textit{e.g.~}}
\nc{\app}{\approx}
\nc{\hf}{\frac{1}{2}}
\nc{\discuss}{\textcolor{red}{Add discussion here!}}
\nc{\red}[1]{\textcolor{red}{#1}}
\nc{\hp}{h_+} 
\nc{\hc}{h_{\times}} 
\nc{\mbh}{M_{\rm BH}}
\nc{\mdisk}{M_{\mathrm{disk}}}
\nc{\rdisk}{r_{\mathrm{disk}}}
\nc{\mc}{M_{\mathrm{c}}}

\begin{document}
	
\title{Whispers from the Early Universe: The Ringdown of Primordial Black Holes}


\author{Chen Yuan}
\email{chenyuan@tecnico.ulisboa.pt}
\affiliation{CENTRA, Departamento de Física, Instituto Superior Técnico – IST, Universidade de Lisboa – UL, Avenida Rovisco Pais 1, 1049–001 Lisboa, Portugal}

\author{Zhen Zhong}
\email{zhen.zhong@tecnico.ulisboa.pt}
\affiliation{CENTRA, Departamento de Física, Instituto Superior Técnico – IST, Universidade de Lisboa – UL, Avenida Rovisco Pais 1, 1049–001 Lisboa, Portugal}

\author{Qing-Guo Huang}
\email{huangqg@itp.ac.cn}
\affiliation{CAS Key Laboratory of Theoretical Physics,
	Institute of Theoretical Physics, Chinese Academy of Sciences,
	Beijing 100190, China}
\affiliation{School of Physical Sciences,
	University of Chinese Academy of Sciences,
	No. 19A Yuquan Road, Beijing 100049, China}
\affiliation{School of Fundamental Physics and Mathematical Sciences
Hangzhou Institute for Advanced Study, UCAS, Hangzhou 310024, China}


\date{\today}

\begin{abstract}
We investigate the stochastic gravitational wave background (SGWB) generated by the ringdown phase of primordial black holes (PBHs) formed in the early universe. As the ringdown signal is independent of the PBH formation mechanism, the resulting SGWB offers a model-independent probe of PBHs. We numerically compute the ringdown waveform and derive the corresponding SGWB. We show that such a signal could be detected by future pulsar timing arrays (PTAs) for PBHs heavier than the solar mass. Additionally, we evaluate the SGWB from binary PBH mergers and demonstrate that it lies within the sensitivity bands of next-generation ground-based interferometers such as Cosmic Explorer and Einstein Telescope, suggesting a multi-band observational strategy for probing the PBH dark matter scenario.

\end{abstract}

\maketitle
\noindent {\bf \em Introduction.} 
The detection of gravitational wave (GW) events from the coalescence of binary black holes and neutron stars by the LIGO and Virgo collaborations has initiated the era of GW astronomy \cite{LIGOScientific:2016emj, LIGOScientific:2016vbw, LIGOScientific:2016vlm}. These observations have renewed interest in primordial black holes (PBHs) as potential constituents of dark matter (DM) since PBH could provide an explanation to the source of these GW events \cite{Sasaki:2016jop, Chen:2018czv, Raidal:2018bbj, DeLuca:2020qqa, Hall:2020daa, Bhagwat:2020bzh, Hutsi:2020sol, Wong:2020yig, DeLuca:2021wjr, Bavera:2021wmw, Franciolini:2021tla, Chen:2021nxo, Chen:2024dxh}.

PBHs are hypothesized to have formed in the early Universe through mechanisms such as the gravitational collapse of overdensed regions generated by enhanced curvature perturbations \cite{Zeldovich:1967lct,Hawking:1971ei,Carr:1974nx}, first-order phase transitions \cite{Hawking:1982ga,Kodama:1982sf}, or the collapse of cosmic strings \cite{Kibble:1976sj,Vilenkin:1984ib,Hindmarsh:1994re,Vilenkin:2000jqa}.
See also \cite{Deng:2016vzb,Saini:2017tsz,Barroso:2024cgg} for different PBH mechanisms.
As such, a wide range of cosmological scenarios predict different stochastic gravitational wave backgrounds (SGWBs) associated with the formation of PBHs. Notable examples include scalar-induced gravitational waves (SIGWs) sourced by large primordial curvature perturbations \cite{Yuan:2021qgz,Domenech:2021ztg}, and GW from first-order phase transitions \cite{Caprini:2015zlo} and cosmic strings \cite{Buchmuller:2021mbb}. See \cite{Khlopov:2008qy,Sasaki:2018dmp,Carr:2020gox,Carr:2020xqk,Green:2020jor,LISACosmologyWorkingGroup:2023njw,Byrnes:2025tji} for reviews of PBHs.

However, all of these GW backgrounds are inherently model-dependent, relying on specific PBH formation mechanisms. Such a fact motivates us to investigate the model-independent GW emission from PBHs.
When PBHs form, they inevitably emit GWs as the spacetime relaxes to a stationary Kerr or Schwarzschild geometry, producing a characteristic quasinormal mode (QNM) signature \cite{Berti:2009kk,Berti:2025hly}.
The ringdown signal is an inevitable byproduct of the black hole formation. As a result, the ringdown SGWB is a potential tool to hunt for PBHs in a model-independent way.

In this work, we compute the SGWB arising from PBH ringdown in the early Universe. To obtain a multi-band signature of the PBHs, we also calculate the SGWB from binary PBH coalescences. Finally, we discuss the astrophysical implications with future GW detectors.

\noindent {\bf \em SGWB from PBH Ringdown.} 
The ringdown waveform, described by the first-order perturbation of one of the Weyl scalars $\Psi_4^{(1)}$, is modeled as a superposition of QNMs, which are found as the eigenmode solutions to the Teukolsky equation with spin weight $s = -2$~\cite{Teukolsky:1973ha,Teukolsky:1972my}. Each QNM has the separable form
\begin{equation}
\psi_{-2}(t, r, \theta, \phi) = 
S(\theta) \, R(r) \, e^{-i \omega t + i m \phi} \,,
\end{equation}  
where $\Psi_4^{(1)} = \psi_{-2}(t, r, \theta, \phi)(r - i a \cos \theta)^{-4}$ and $\omega$ is the complex QNM frequency. To create a signal with a distinct beginning and to mimic propagation along the light cone, we add a cutoff on the waveform by multiplying $\Psi_4$ with a Heaviside theta function (see e.g., \cite{Zhong:2024ysg} for details).

\begin{figure}
	\centering
	\includegraphics[width = 0.45\textwidth]{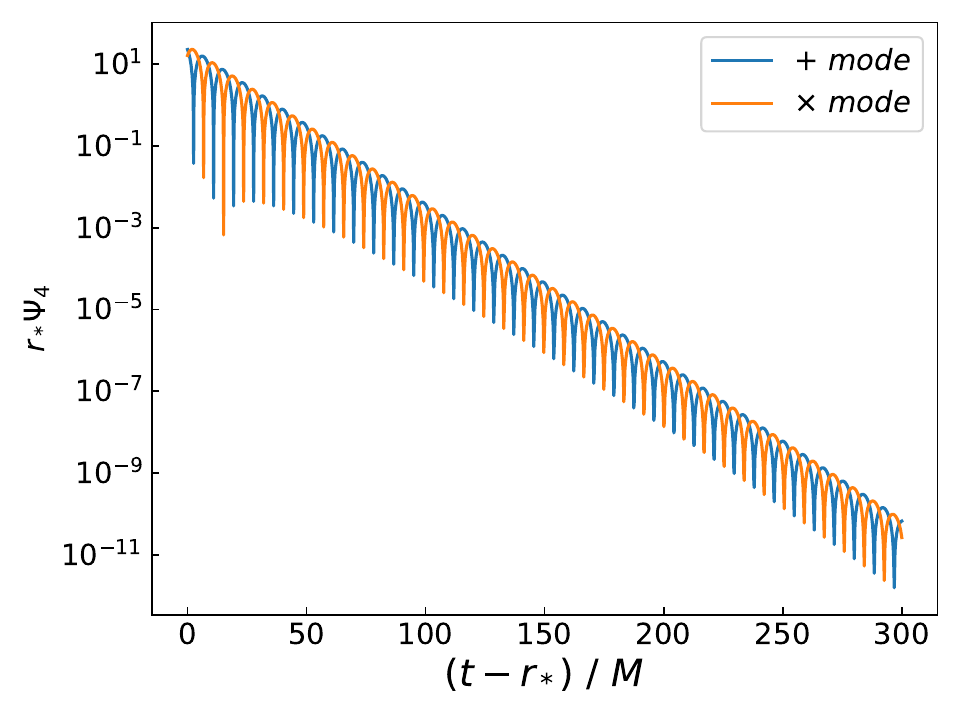}
    \caption{\label{ringdown} The ringdown signal for a non-spinning black hole, extracted at $r_*=800M$. This result corresponds to fundamental mode $M\omega=0.3737-0.08896i$ with $l = m = 2$. }
\end{figure}

We focus on the fundamental mode $M\omega=0.3737-0.08896i$ with $l = m = 2$ for non-spinning black holes. The numeric waveform of the ringdown stage is shown in Fig.~\ref{ringdown}. Given the result of $\Psi_4$, the two GW polarizations, $h_+$ and $h_{\times}$, can be obtained by double time integration
\begin{equation}
\Psi_4=\frac{1}{2}\left(\frac{\partial^2 h_{+}}{\partial t^2}-i \frac{\partial^2 h_{\times}}{\partial t^2}\right)\,.
\end{equation}

The SGWB arises from the superposition of uncorrelated GWs coming from all directions in the sky, and its dimensionless energy spectrum is given by \cite{Phinney:2001di}
\begin{equation}
    \Omega_{\mathrm{GW}}(f)=\frac{f}{\rho_c}\int dz~ n(z) {\frac{dE}{df_s}}\Bigg|_{f_s=f(1+z)},
\end{equation}
where the number of events per comoving volume within $[z,z+dz]$ is represented by $n(z)$ and $dE/df_s$ stands for the energy spectrum of ringdown in the source frame, namely
\begin{equation}
\frac{\mathrm{d} E}{\mathrm{d}f_s} = \int\mathrm{d} \Omega\frac{\pi^2 c^3 }{2G} r^2f_s^2\left( |\tilde{h}_+(f_s, \theta, \phi)|^2 + |\tilde{h}_\times(f_s, \theta, \phi)|^2 \right),
\end{equation}
In this study, we focus on monochromatic PBHs so that
\begin{equation}\label{nz}
    n(z) = {f_{\mathrm{pbh}}  \over M}\Omega_{\mathrm{cdm}} \rho_c  \frac{dt}{dz}\delta(t-t_{\mathrm{form}}),
\end{equation}
where $dt/dz$ is the derivative of the lookback time with respect to redshift, given by
\begin{equation}
    \frac{dt}{dz} = \frac{1}{H_0 \sqrt{\Delta} (1 + z)}.
\end{equation}
Here $H_0$ is the Hubble value by today \cite{Planck:2018vyg}, $\Delta = \Omega_{r}(1+z)^{4}+\Omega_{m}(1+z)^{3}+\Omega_{\Lambda}$, $\Omega_{r}$, $\Omega_m$ and $\Omega_\Lambda$ are density parameters for radiation, matter and dark energy respectively \cite{Planck:2018vyg}. The PBH abundance is characterized by $f_{\mathrm{pbh}} \equiv \Omega_{\mathrm{pbh}} / \Omega_{\mathrm{cdm}}$ and $\Omega_{\mathrm{cdm}}=\rho_{\mathrm{cdm}}/\rho_c$. The PBH mass is related to the formation time, $t_{\mathrm{form}}$, through \cite{Carr:1974nx,Carr:1975qj}
\begin{equation}
    M\simeq 2\times10^5M_{\odot}\left(\frac{t_{\mathrm{form}}}{1s}\right).
\end{equation}
To further quantify the SGWB, we also introduce a parameter, $\epsilon$, to rescale the GW spectrum such that the fraction of PBH mass converted into GWs during the ringdown stage is described as
\begin{equation}
    \int_{0}^{+\infty}\frac{dE}{df}df=\epsilon M.
\end{equation}
As a benchmark example, the total radiated energy during the ringdown stage is $\sim 3\%$ for binary black hole merger \cite{Berti:2007fi}.

The result of ringdown SGWB is demonstrated in Fig.~\ref{SGWB}, assuming $\epsilon f_{\mathrm{pbh}}= 10^{-6}$. 
Also shown are the power-law integrated sensitivity curves of future 30-year observations from IPTA and SKA. We also show the sensitivity curves of ``$\ddot{P}$ analysis'' for these two PTAs \cite{DeRocco:2022irl}. It can be seen that the ringdown SGWB could fall within the detectable range of future PTAs for PBHs with masses $M\gtrsim M_{\odot}$.

\begin{figure*}
	\centering
	\includegraphics[width = 0.45\textwidth]{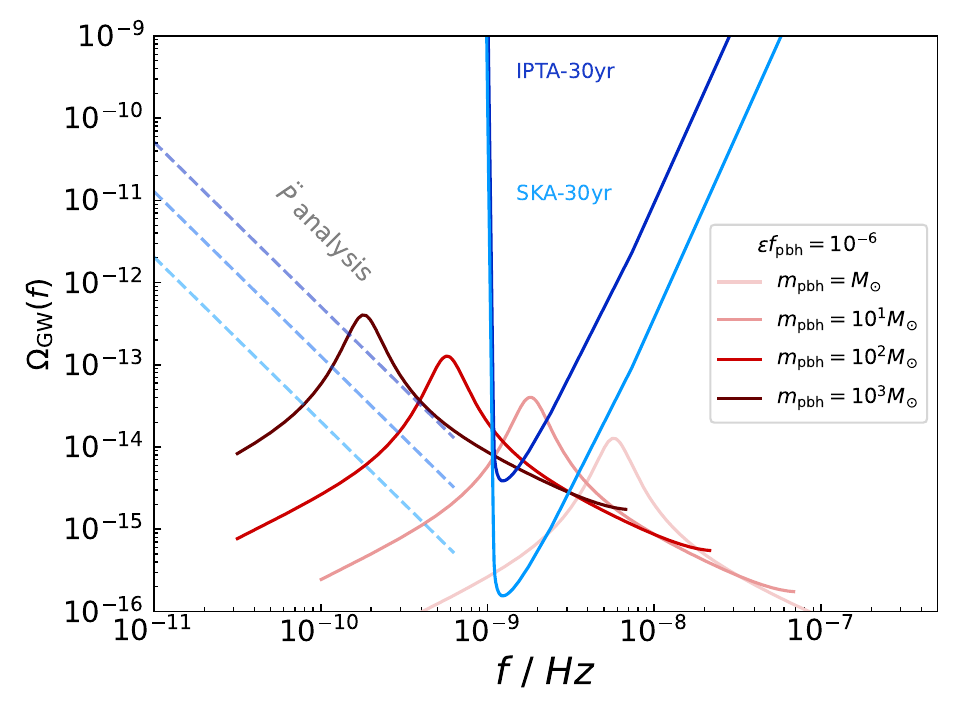}
    \includegraphics[width = 0.45\textwidth]{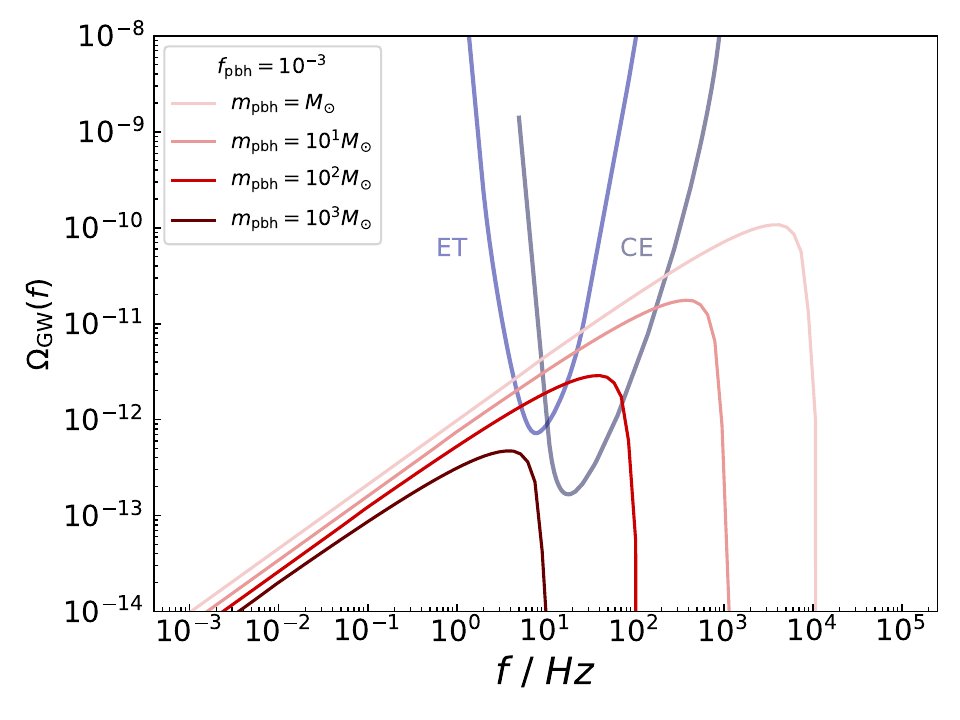}
    \caption{\label{SGWB} Left panel: The SGWB from PBH ringdown, together with the future pulsar timing array sensitivity curves. Right panel: The SGWB from binary PBH merger together with sensitivity curves of CE and ET.}
\end{figure*}

\noindent {\bf \em SGWB form binary PBH coalescences.} 
The SGWB originating from binary PBH mergers across the Universe can be evaluated following ~\cite{Phinney:2001di,Regimbau:2008nj,Zhu:2011bd,Zhu:2012xw}:
\begin{equation}\label{SGWB}
    \Omega_{\mathrm{GW}}(f) = \frac{f}{\rho_c} \int d M_1 d M_2 dz \, \frac{dt}{dz} \, R(z, m_1, m_2) \frac{dE_s}{df_s},
\end{equation}
The energy spectrum of a single coalescence event in the source frame, $dE_s/df_s$, can be found in Refs.~\cite{Cutler:1993vq,Chernoff:1993th,Zhu:2011bd}.

For general PBH mass functions, the merger rate density is expressed as~\cite{Hutsi:2020sol,Franciolini:2022tfm}
\begin{equation}\label{R12}
\begin{aligned}
&R_{12}(z, m_1, m_2)(t \mid \vec{\theta}) = \frac{1.6 \times 10^6}{\mathrm{Gpc}^3 \, \mathrm{yr}} \,
f_{\mathrm{pbh}}^{\frac{53}{37}} \left( \frac{t(z)}{t_0} \right)^{-\frac{34}{37}} 
\eta^{-\frac{34}{37}} \\
&\quad \times \left( \frac{M}{M_\odot} \right)^{-\frac{32}{37}} S\left[M, f_{\mathrm{pbh}}, P(m), z\right] 
P(m_1) P(m_2),
\end{aligned}
\end{equation}
where $M = m_1 + m_2$ is the total mass, $\eta = m_1 m_2 / M^2$ is the symmetric mass ratio, and $t_0$ denotes the present age of the Universe. The symbol $\vec{\theta}$ collectively denotes the parameters of the PBH mass function, normalized via $\int P(m) \, \mathrm{d}m = 1$. The suppression factor $S$ is detailed in~\cite{Hutsi:2020sol,Franciolini:2022tfm}.

We also incorporate the contribution from PBH mergers arising in three-body formation channels, following~\cite{Vaskonen:2019jpv,Raidal:2024bmm}:
\begin{equation}
    \begin{aligned}
&R_{12,3}(z, m_1, m_2)(t \mid \vec{\theta}) \approx \frac{7.9 \times 10^4}{\mathrm{Gpc}^3 \, \mathrm{yr}} \left( \frac{t}{t_0} \right)^{\frac{\gamma}{7} - 1} f_{\mathrm{pbh}}^{\frac{144 \gamma}{259} + \frac{47}{37}} \\
&\qquad \times \left[ \frac{\langle m \rangle}{M_\odot} \right]^{\frac{5 \gamma - 32}{37}} \left( \frac{M}{2 \langle m \rangle} \right)^{\frac{179 \gamma}{259} - \frac{2122}{333}} (4 \eta)^{-\frac{3 \gamma}{7} - 1} \\
&\qquad \times \mathcal{K} \frac{e^{-3.2(\gamma - 1)} \gamma}{28/9 - \gamma} \, \overline{\mathcal{F}}(m_1, m_2) \, P(m_1) \, P(m_2),
\end{aligned}
\end{equation}
where $\gamma$ and $\mathcal{K}$ quantify the angular momentum distribution and binary hardening due to PBH encounters. Following numerical simulations~\cite{Raidal:2018bbj}, we adopt $\gamma = 1$ and $\mathcal{K} = 4$. The $\overline{\mathcal{F}}$ function is defined by
\begin{equation}
\begin{aligned}
\overline{\mathcal{F}}(m_1, m_2) &\equiv \int_{m \leq m_1, m_2} \mathrm{d}m \, P(m) \frac{\langle m \rangle}{m} \\
&\quad \times \left[ 2 \mathcal{F}(m_1, m_2, m) + \mathcal{F}(m, m_1, m_2) \right],
\end{aligned}
\end{equation}
with the three-body enhancement factor $\mathcal{F}$ given by
\begin{equation}
\begin{aligned}
\mathcal{F}(m_1, m_2, m_3) &= m_1^{5/3} m_2^{5/3} m_3^{7/9}
\left( \frac{m_1 + m_2}{2} \right)^{4/9} \\
&\quad \times \left( \frac{m_1 + m_2 + m_3}{3} \right)^{2/9}
\langle m \rangle^{-43/9}.
\end{aligned}
\end{equation}

The SGWB generated by binary PBH coalescences is shown in Fig.~\ref{SGWB}. We find that PBH binaries with masses ranging from solar mass to intermediate-mass scales can produce a SGWB that falls within the sensitivity range of future ground-based GW detectors such as Cosmic Explorer (CE) \cite{LIGOScientific:2016wof} and Einstein Telescope (ET) \cite{ET:2019dnz,Punturo:2010zz}. See also \cite{Ackley:2020atn,Capdevilla:2025ivz}.
This enables a promising multi-band strategy: ringdown signals could be probed by PTAs in the sub-nanohertz to nanohertz band, while the corresponding merger signals could be detected at higher frequencies by ground-based detectors.

\noindent {\bf \em Constraints on PBH abundance.} 
To quantify the observational constraints on $f_{\mathrm{pbh}}$ by future GW detectors, we estimate the signal-to-noise ratio (SNR) of the ringdown SGWB by PTAs, namely \cite{Allen:1997ad,Thrane:2013oya}
\begin{equation}
\rho^{2} \simeq  2T \sum_{I,J}^{N}\zeta_{IJ} \int \mathrm{d} f \frac{S_{h}(f)^{2}}{P_{n}(f)^{2}/R(f)^{2}},
\end{equation}
where $R(f)$ and $P_n(f)$ are the detector response and the noise power spectral density \cite{Thrane:2013oya}. The strain power spectral density, $S_h(f)$, is related to the energy density of SGWB through $S_h(f) = 3H_0^2\Omega_{\mathrm{GW}}(f)/(2\pi^2 f^3)$. The normalized Hellings and Downs coefficient for pulsars $I$ and $J$ is denoted by $\zeta_{IJ}$. For $N$ pulsars distributed homogeneously in the sky, we have
\begin{equation}
    \sum_{I,J}^{N}\zeta_{IJ}\simeq \frac{N(N-1)}{96}.
\end{equation}
For IPTA/SKA, we assume $200$ pulsars and $14$ days of cadence for both PTAs, $100$ns and $20$ns timing accuracy respectively. As a conservative estimate, we do not include the ``$\ddot{P}$ analysis'' when computing the SNR.

\begin{figure}
	\centering
	\includegraphics[width = 0.45\textwidth]{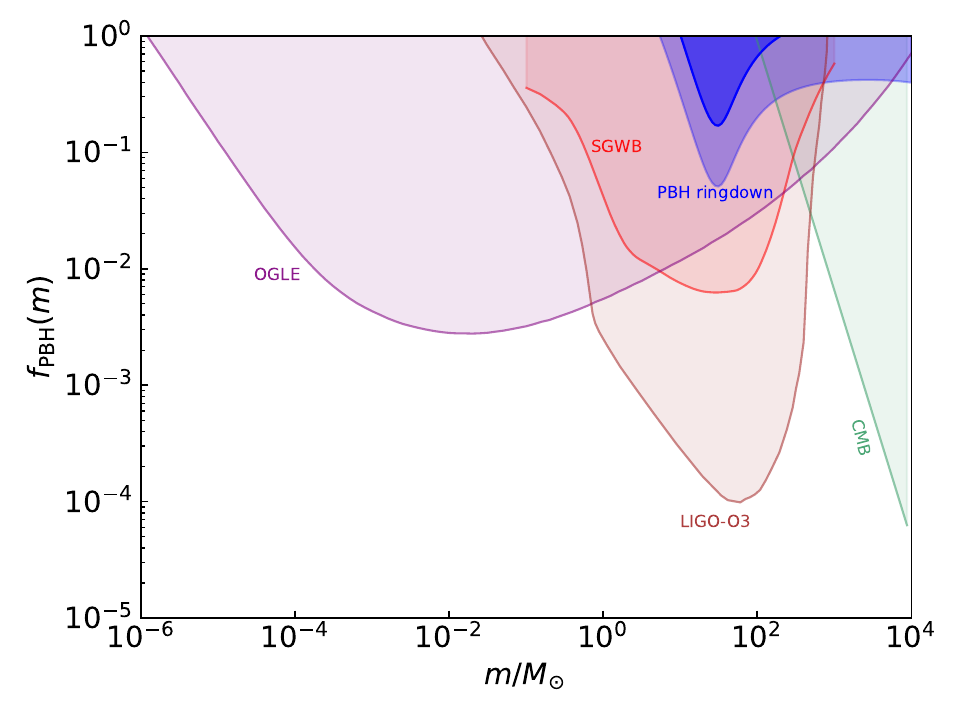}
    \caption{\label{fpbh} Constraints on $f_{\mathrm{fpbh}}$ for monochromatic PBHs. For the ringdown SGWB, we show the expected constraints by SKA, assuming $\epsilon=3\%$ (dark blue) and $\epsilon=10\%$ (light blue) and a threshold SNR, $\rho=5$, for claiming the detection of SGWB. The parameters for SKA are: $200$ pulsars, $20$ns of timing accuracy, $14$ days of cadence and $30$yr of observations.}
\end{figure}

Fixing $\rho=5$ as a threshold SNR for claiming the detection of a SGWB and $3\%$ of the PBH mass converts into GWs, we compute the upper limit of $f_{\mathrm{pbh}}$. The results are illustrated in Fig.~\ref{fpbh}, together with constraints from the OGLE microlensing \cite{Mroz:2024mse,Mroz:2024wag}, the null detection of SGWB from binary PBHs by LIGO-Virgo-KAGRA \cite{Nitz:2022ltl,Boybeyi:2024mhp}, the LIGO-O3 results by assuming all the O3 events have astrophysical origin \cite{Hutsi:2020sol,Andres-Carcasona:2024wqk}, accretion constraints of CMB \cite{Ali-Haimoud:2016mbv,Blum:2016cjs,Horowitz:2016lib,Chen:2016pud,Poulin:2017bwe}. We find that under the above assumptions, IPTA cannot place meaningful constraints on $f_{\mathrm{pbh}}$, unless future improvements in timing precision, number of pulsars, or observation time are achieved. In contrast, SKA will be able to constrain PBHs with masses from solar to intermediate scales down to $f_{\mathrm{pbh}}\sim \mathcal{O}(0.1)$, providing an alternative way to search for or constrain the PBH population.

\noindent {\bf \em Discussion and Conclusion} 
By numerically modeling the ringdown waveform, we obtained a detailed prediction for the SGWB during PBH formation, which is independent of the PBH formation channels. We focused on monochromatic PBHs, and demonstrated that the corresponding ringdown SGWB falls in the nanohertz band, could be detectable by future pulsar timing arrays such as IPTA/SKA. Note that the peak frequency of the ringdown spectrum is proportional to $M$ and multiply with the red-shift factor (see e.g., \cite{DeLuca:2025uov}). As such, in the standard PBH formation channels where PBHs are generated from the collapse of overdensities, the ringdown background would be buried beneath the scalar-induced gravitational waves whose frequency is also proportional to $M$ times the red-shift factor (see Eq.~(4.12) in \cite{Domenech:2023fuz}). However, for other formation channels such as domain walls, the peak frequency of GWs would be different from that of the ringdown signal \cite{Li:2024psa}. Hence, the ringdown signal might be resolvable in such a case.

In addition, we computed the SGWB generated by binary PBH mergers using both two-body and three-body formation channels. We showed that these merger-induced signals, particularly for solar mass PBHs to intermediate PBHs, fall within the sensitivity bands of next-generation ground-based gravitational wave detectors such as CE/ET. Therefore, future GW observations in both low and high frequency regimes may provide complementary constraints on the PBH population.

Looking forward, our study can be extended in several directions. First, the amplitude of the ringdown signal depends on the excitation efficiency of QNMs, which can be more accurately determined through full numerical relativity simulations of PBH formation in specific cosmological scenarios. Second, our analysis assumes a monochromatic mass function; incorporating extended mass distributions and PBH spins would improve the SGWB prediction. Moreover, including potential environmental effects, such as the presence of surrounding matter or dark radiation, may slightly shift the QNM frequencies and affect the detectability of the signal.
Finally, although our results show that the ringdown background of PBHs heavier than solar mass could fall in the PTA band, we remind the readers that we adopt an optimistic PTA configuration. A more precise PTA data analysis is to consider the SGWB that the PTAs have already observed as a foreground, see e.g., \cite{Babak:2024yhu} for details.

Despite these limitations, our work highlights the power of ringdown SGWB to probe the existence and properties of PBHs across a broad mass range. The unavoidable nature of the ringdown SGWB makes it a unique probe of PBHs, which is independent of the PBH formation channels.

\noindent {\bf \em Note added.} 
While we were finalizing this work, \cite{DeLuca:2025uov} appeared on arXiv, which also studied the SGWB from PBH ringdown. Compared to their analysis, which focuses on extremely massive PBHs and their potential detectability through CMB B-modes, our work investigates the SGWB from stellar to intermediate-mass PBHs and their observability via PTAs. Furthermore, we numerically compute the ringdown waveform and additionally include the SGWB contribution from binary PBH mergers, offering a complementary multi-band perspective. We have also verified that our results, the peak frequency of the SGWB for instance, are consistent with those in \cite{DeLuca:2025uov}. Notably, their approach of rescaling the ringdown amplitude, $A$, is equivalent to our parametrization using the energy fraction $\epsilon$ converted GWs during the ringdown phase.

\section*{Acknowledgments}
CY acknowledges financial support provided under the European Union’s H2020 ERC Advanced Grant “Black holes: gravitational engines of discovery” grant agreement no. Gravitas–101052587. 
Views and opinions expressed are however those of the author only and do not necessarily reflect those of the European Union or the European Research Council. Neither the European Union nor the granting authority can be held responsible for them.
This project has received funding from the European Union's Horizon 2020 research and innovation programme under the Marie Sklodowska-Curie grant agreement No 101007855 and No 101131233. QGH is supported by the National Key Research and Development Program of China Grant No.2020YFC2201502 and the grants from NSFC (Grant No.~12475065, 12447101). Z.Z.\ acknowledges financial support from China Scholarship Council (No.~202106040037).

\bibliography{ref}
\end{document}